\documentclass[preprint]{aastex}

\input{epsf}

\newcommand{\gsim}{\mathrel{\hbox{\rlap{\hbox{\lower4pt\hbox{$\sim$}}}\hbox{$>$}}}}
\newcommand{\lsim}{\mathrel{\hbox{\rlap{\hbox{\lower4pt\hbox{$\sim$}}}\hbox{$<$}}}}

\shortauthors{Hoyle, Vogeley \& Gott} 
\shorttitle{Two-Dimensional Topology of the 2dFGRS}

\begin{document}

\title{Two-Dimensional Topology of the 2dF Galaxy Redshift Survey}

\author{Fiona Hoyle$^1$, Michael S. Vogeley$^1$ \& J. Richard Gott III$^2$ \\ 
1. Department of Physics, Drexel University, 
3141 Chestnut Street, Philadelphia, PA 19104 \\ 
2. Princeton University Observatory, Princeton, NJ 08544 \\
\email{hoyle@venus.physics.drexel.edu, vogeley@drexel.edu, jrg@astro.princeton.edu }
}

\begin{abstract}

We study the topology of the publicly available data released by the 2
degree Field Galaxy Redshift Survey team (2dFGRS). The 2dFGRS data
contains over 100,000 galaxy redshifts with a magnitude limit of
b$_{\rm J}$=19.45 and is the largest such survey to date. The data lie
over a wide range of right ascension (75$^{\circ}$ strips) but only
within a narrow range of declination (10$^{\circ}$ and 15$^{\circ}$
strips). This allows measurements of the two-dimensional genus to be
made.

We find the genus curves of the NGP and SGP are slightly
different. The NGP displays a slight meatball shift topology, whereas
the SGP displays a bubble like topology. The current SGP data also have
a slightly higher genus amplitude. In both cases, a slight excess
of overdense regions are found over underdense regions. We assess the
significance of these features using mock catalogs drawn from the
Virgo Consortium's Hubble Volume $\Lambda$CDM $z$=0 simulation. We
find that differences between the NGP and SGP genus curves are only
significant at the 1$\sigma$ level. The average genus curve of the
2dFGRS agrees well with that extracted from the $\Lambda$CDM mock
catalogs.

We also use the simulations to assess how the current incompleteness
of the Survey (the strips are not completely filled in) affects the
measurement of the genus and find that we are not sensitive to the
geometry; there are enough data in the current sample to trace the
isolated high and low density regions.

We compare the amplitude of the 2dFGRS genus curve to the amplitude of
a Gaussian random field with the same power spectrum as the 2dFGRS and
find, contradictory to results for the 3D genus of other samples, that the
amplitude of the GRF genus curve is slightly lower than that of the
2dFGRS. This could be due to a a feature in the current data set or
the 2D genus may not be as sensitive as the 3D genus to non-linear
clustering due to the averaging over the thickness of the slice in 2D.

\end{abstract}


\section{Introduction}

Matching the large scale structure of the universe remains one of the
most important constraints for models of structure formation. Models
have been primarily compared to observations using statistics such as
the correlation function, power spectrum and counts-in-cells. These
two-point statistics have shown that galaxies are indeed clustered and
have allowed us to reject models, such as the standard cold dark
matter model (SCDM) with $\Omega_{\rm m}=1, \Gamma = 0.5$, as it 
does not have enough power on large scales. Similar
constraints from the cosmic microwave background measurements,
supernova Ia experiments, large scale structure and clusters (see
Bahcall et al. 1999 for a recent review) have led to the development
of a consensus flat cosmological constant-dominated CDM model.

All currently examined CDM models begin with Gaussian random phase
initial conditions. Inflation predicts that the seeds for structure
formation should derive from a Gaussian random phase distribution
(Bardeen, Steinhardt \& Turner 1983). This can be tested using
topological statistics such as the genus statistic (Gott et
al. 1986). The topology of the large scale structure is invariant
during the linear growth of structure, thus, after appropriate
smoothing on large scales, the topology of the present galaxy
distribution can be related to that of the initial density field. This
allows a test of the random phase hypothesis as any deviation of the
measured topology might be evidence for non-Gaussian initial
conditions. On relatively smaller scales, the topology of the smoothed
galaxy density quantifies the degree of non-linear evolution and/or
biasing of galaxy formation with respect to the mass density at the
present epoch.

The genus statistic has been measured from many different galaxy
redshift surveys. Gott et al. (1989) applied the 3D genus statistic to
small samples of galaxies. It has since been applied to
subsequently larger surveys; the 3D genus of the SSRS was measured by
Park, Gott \& da Costa (1992), Moore et al. (1992) applied it to the
QDOT Survey, Rhoads et al. (1994) analyzed Abell Clusters, Vogeley et
al. (1994) analyzed the CfA survey and Canavezes et al. (1998) analyzed the
PSCz Survey. In 2D, the method has been applied to the CfA Survey
(Park et al. 1993) and the LCRS survey (Colley 1997). The 2D genus of
the microwave background has also been measured (Colley, Gott \& Park
1996). These papers have all concluded that the genus curve obtained
was consistent with the universe having Gaussian initial conditions
but in some cases, results were limited by the size of the
surveys. Some evidence for departure from Gaussianity was found as a
result of non-linear gravitational evolution and/or biasing of
galaxies as compared to the mass.

In this paper, we estimate the genus of the largest galaxy redshift
survey that is publicly available, the 2 degree Field Galaxy Redshift
Survey (2dFGRS). In section \ref{sec:survey} we describe the survey in
more detail. In section \ref{sec:genus} we describe the genus
statistic and in section \ref{sec:res} we present our results. We draw
conclusions in section \ref{sec:conc}.

\section{The 2dFGRS}
\label{sec:survey}
The 2dFGRS is an optical spectroscopic survey of objects brighter than
b$_{\rm J}$=19.45 selected from the APM Galaxy Survey (Maddox et
al. 1990a, b). The 2dFGRS survey is divided into two main
regions, with additional random fields observed to improve the angular
coverage for statistics such as the power spectrum. For this analysis
we do not use the random fields. The two areas of interest are the
South Galactic Pole (SGP) region, which covers the region $
325^{\circ} < \alpha < 52.5^{\circ}$, $-37.5^{\circ} < \delta <
-22.5^{\circ}$ and a region which is close to the North Galactic Pole
(NGP), $ 147.5^{\circ} < \alpha < 222.5^{\circ}$, $-7.5^{\circ} <
\delta < 2.5^{\circ}$.

The data that we analyze are from the public release that was
distributed to the community in June 2001 (Colless et al. 2001 and
references therein). 102,426 galaxy redshifts were contained in the
data release, making this the largest redshift survey that is publicly
available. As the survey is not finished, the angular selection
function is complicated. To match this when we construct random
catalogs, required when estimating the genus, we use the software
developed by the 2dFGRS team and distributed as part of the data
release \footnote{for the data release products and catalogs see {\tt
www.mso.anu.edu.au/2dFGRS/}}. For any given coordinates ($\alpha, \delta$),
the expected probability of a galaxy being contained in
the 2dFGRS survey region is returned.

We construct a volume-limited sample from the 2dFGRS in order to
obtain a radial selection function that is uniform, thus the only
variation in the space density of galaxies with radial distance is due
to clustering. This means no weighting scheme is required when
constructing the density field but it does restricts the number of
objects in a volume-limited sample extracted from this data. We select
a volume limit of $z_{\rm max}$=0.138. This value maintains the
maximum number of galaxies in the sample. We adopt the global
$k$-correction + evolution correction found for b$_{\rm J}$ selected
galaxies in the ESO Slice Project (Zucca et al. 1997) and adopted by
the 2dFGRS team (Norberg et al. 2001) and use
\begin{equation}
k + e = \frac{0.03 z}{(0.01 + z^4)}.
\end{equation} 
Following the 2dFGRS team and to maintain consistency with the
simulation we will describe in Section \ref{sec:res}, we adopt a
$\Omega_{\rm m}=0.3, \Omega_{\Lambda}=0.7$ cosmology when converting
redshift into comoving distances. For this cosmology, the redshift
limit of $z_{\rm max}$=0.138 corresponds to a comoving distance of
398$h^{-1}$Mpc.

\begin{figure} 
\begin{centering}
\hspace{-1cm}{\epsfxsize=6truecm \epsfysize=2truecm \epsfbox[0 50 550 200]{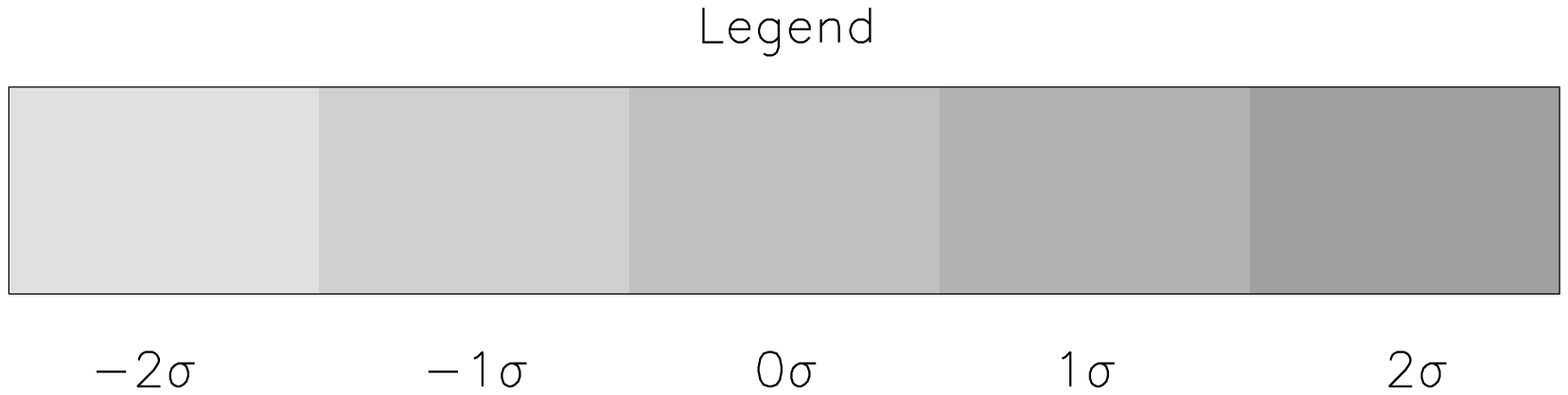}} 
\begin{tabular}{cc}
{\epsfxsize=6truecm \epsfysize=6truecm \epsfbox[100 100 500 500]{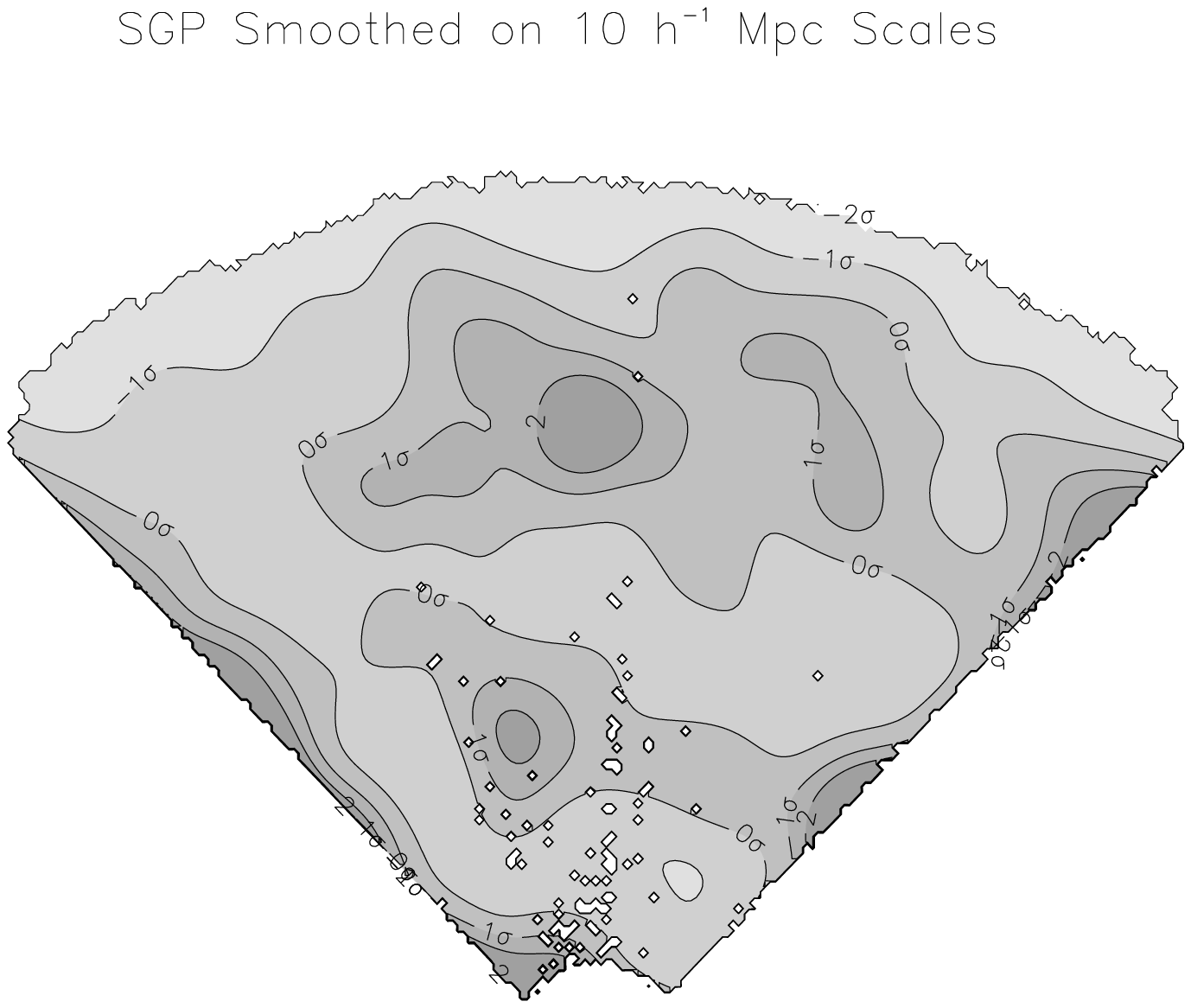}} &
{\epsfxsize=6truecm \epsfysize=6truecm \epsfbox[100 100 500 500]{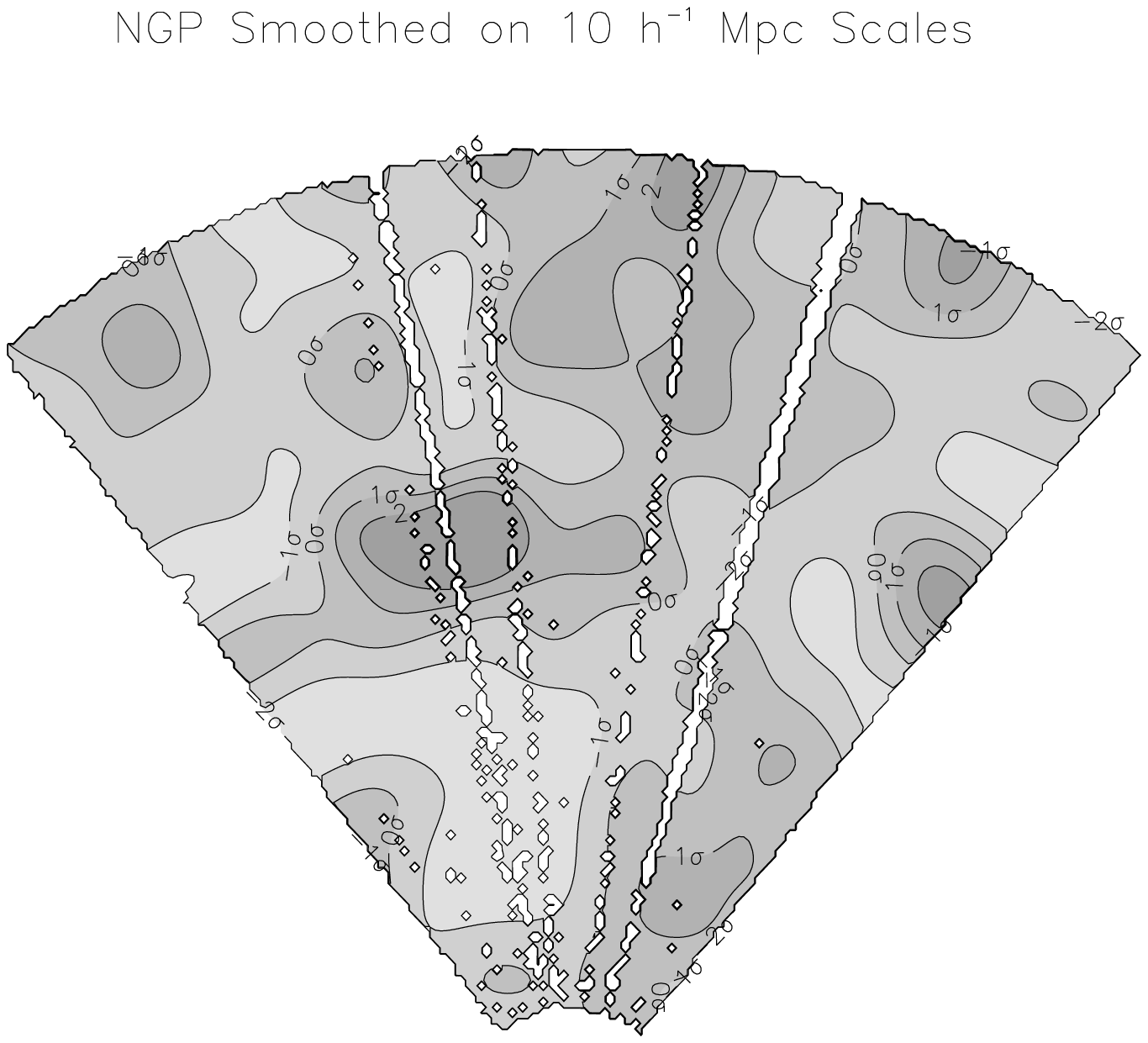}} \\
{\epsfxsize=6truecm \epsfysize=6truecm \epsfbox[100 250 500 750]{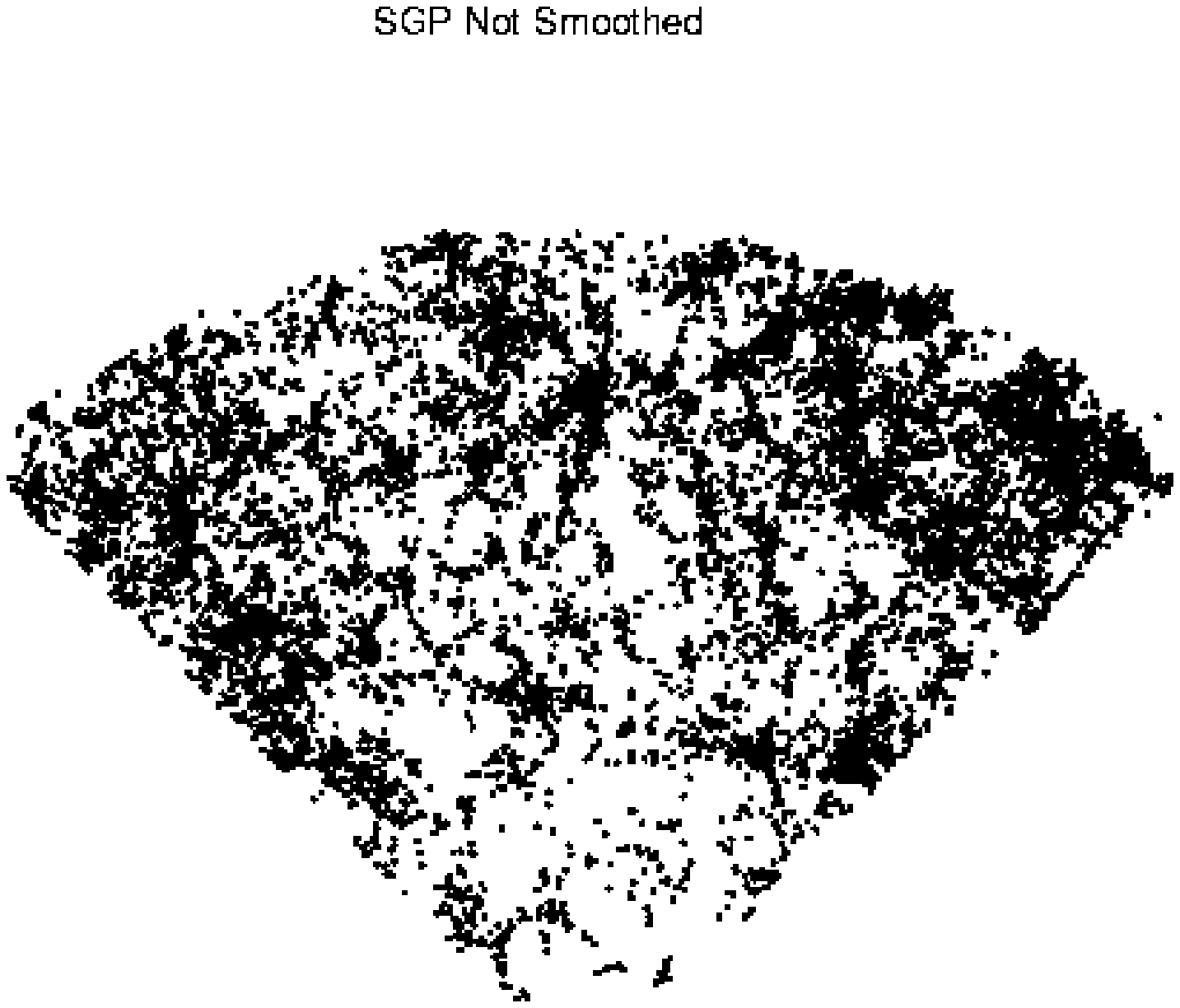}} &
{\epsfxsize=6truecm \epsfysize=6truecm \epsfbox[100 250 500 750]{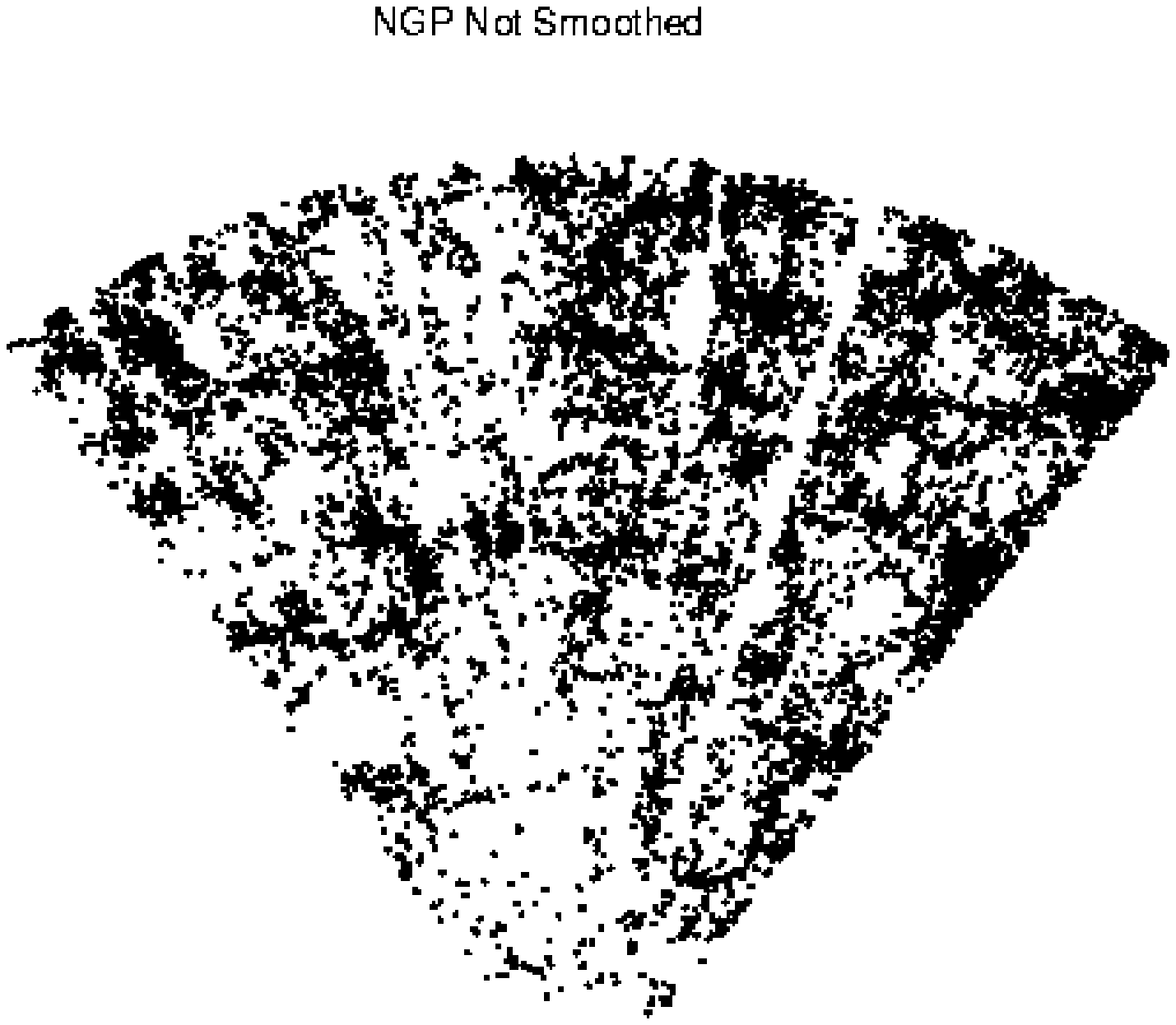}} \\
\end{tabular}
\caption{Smoothed and unsmoothed contour maps of the density of
objects in the SGP and NGP slices. The darker the contour, the denser
the region. In the smoothed plots, the darkest contours are 2$\sigma$
over-dense regions and the lightest contours are the 2$\sigma$
under-dense regions.
The 2D smoothing kernel is a Gaussian with dispersion $\lambda =10h^{-1}$Mpc.
Small differences in the qualitative appearance of the smoothed and unsmoothed maps are due to incompleteness of the survey. This incompleteness and the varying thickness of the slices with redshifts are accounted for in the smoothed maps.
 }
\label{fig:smooth}
\end{centering}
\end{figure}

In the NGP region we have 11375 galaxies and in the SGP region there
are 12062 galaxies in our volume limited sample. The mean 3D
galaxy-galaxy separation in the volume-limited samples is of order
7$h^{-1}$Mpc. This is then projected down into two dimensions, which
reduces the mean 2D galaxy-galaxy separation. Therefore our
choice of 10$h^{-1}$Mpc for the smoothing length in most of the genus 
calculations (see section
\ref{sec:genus}) is large enough so that the genus signal is not
dominated by Poisson shot noise.

\subsection{The Simulation}
\label{sec:sims}

To test the robustness of our statistical methods, estimate
uncertainties in our measurement, and to test the currently best
fitting variant of CDM, we use mock surveys drawn from the Virgo
Consortium's Hubble Volume $z=0$ $\Lambda$CDM simulation. This
simulation contains 1 billion mass particles in a cube with side
3,000$h^{-1}$Mpc, large enough that many independent mock catalogs
can be constructed. The cosmological parameters of the simulation are
$\Omega_{\rm b}=0.04, \Omega_{\rm CDM}=0.26, \Omega_{\Lambda}=0.7,
h=0.7, \Gamma=0.17$, close to the values suggested by median
statistics analysis (Gott et al. 2001).  The clustering amplitude of
the dark matter particles closely matches that of present day,
optically selected galaxies so no biasing has been applied. For more
details see Frenk et al. (2000). This simulation does not
include any gas physics but we generally smooth on large, 10$h^{-1}$Mpc, scales
where the dark matter fluctuations might be expected to trace the
clustering pattern of galaxies.

We construct samples that have the same geometry as the current 2dFGRS
data using the same technique that we use to construct the random
catalog. The only difference is that in the mock catalogs we
sparsely sample the points to match the number density of galaxies in
the 2dFGRS.

\section{The Genus Statistic}
\label{sec:genus}

The genus statistic is a quantitative measure of the topology of
regions bounded by isodensity surfaces. In two-dimensions, that
surface is simply the set of curves that separate high and low density
areas. Here, we present results on the two-dimensional genus statistic
(see Gott et al. 1986; Melott 1989; Melott et al. 1989; Gott et
al. 1990; Park et al. 1992; Colley 1997). Following Melott et
al., we define the 2-D genus to be

\begin{equation}
G_2 = {\rm number \; of \; isolated \; high \; density \; regions \; - \; number \; of \; isolated \; low \; density \; regions.}
\end{equation}

The genus of a contour in a two-dimensional density distribution can also be calculated using the Gauss-Bonet theorem, in the two dimensional form, using
\begin{equation}
G_2 = \frac{\int C dS}{2 \pi}
\end{equation}
where the line integral follows the contour and $C$ is the inverse curvature $r^{-1}$ of the line enclosing a high or low density region. This can be positive or negative depending on whether a high or low density region is enclosed. The genus of a contour enclosing an isolated high density region is positive, where as the genus of a contour enclosing a low density region is negative. This expression allows contributions from contours around regions that do not lie fully within the boundaries of the survey to be included.

For a Gaussian random field, the genus has a simple analytic form. The
genus per unit area is expressed as
\begin{equation}
g_2 = \frac{G_2}{\rm Area} = A \nu e^{- \nu^2/2},
\label{eq:genus}
\end{equation}
where $\nu$ is the threshold value, above which a fraction, $f$, of the area has a higher density
\begin{equation}
f = (2\pi)^{-1/2} \int^{\infty}_{\nu} \exp(-x^2/2) dx
\end{equation}
The value of the amplitude, $A$, depends on the shape of the power
spectrum. Melott et al. (1989) describe the form that $A$ can take in
extreme situations for varying power spectrum indices. i.e. if the
smoothing length is much larger or smaller than the thickness of the
slice. In our case the smoothing length is comparable to the thickness
of the slice and the power spectrum index of the 2dFGRS is close to
$n=-2$, thus neither of the approximations given in Melott et al. are
appropriate here.

The steps to calculating the genus are as follows:
\begin{itemize}
\item Construct a random catalog of points for the NGP and SGP that
have the same angular and radial selection function as the data points
but with much higher number density
\item Take the data points (the NGP and SGP points are considered
separately) and the random points and place each of them on a grid. We
have 256$^2$ cells in each grid. Calculate the density of data and
random points in each cell.
\item Smooth the data and random density field with a Gaussian kernel
\item Divide the smoothed data density field by the smoothed random
density field
\item Mark the cells that lie outside the survey (i.e. the cells with
density=0 in the unsmoothed random density field) with a negative
value. These cells are then ignored by Contour2D
\item Run CONTOUR2D
\end{itemize}

We smooth the data and random point density distributions by
convolving the density field with a Gaussian of the form $e^{-r^2/(2
\lambda)^2}$, where $\lambda$ is the smoothing scale, typically set
here to 10$h^{-1}$Mpc. This value is chosen so that the structures
that remain are in the quasi-linear regime.  Note that this definition
of the Gaussian smoothing length $\lambda$ differs from some earlier
papers on the genus statistic (some earlier papers use $e^{-r^2/{\lambda}^2}$).
The smoothed and unsmoothed NGP and
SGP slices are shown in figure \ref{fig:smooth}. Structures are
clearly able to survive the small gaps in the currently incomplete
data. The random points are used to mark the edge of the survey and to
take out any contribution to the genus from the shape of the survey
(c.f. the window function in power spectra analysis).

CONTOUR2D \footnote{Note that the version of CONTOUR2D published in 
Melott 1989 contains a sign error in line 19 of the OUT subroutine.} 
uses a two-dimensional variant of the angle deficit
algorithm described by Gott et al. (1986). The density contour is
specified by the fractional area contained in the high density region
of the smoothed density field. The program classifies all cells as
high or low density then evaluates the genus of the contour surface by
summing the angle deficits at all vertices that lie on the boundary
between the high and low density regions.

\begin{figure} 
\begin{centering}
\begin{tabular}{c}
{\epsfxsize=8truecm \epsfysize=8truecm \epsfbox[40 120 600 700]{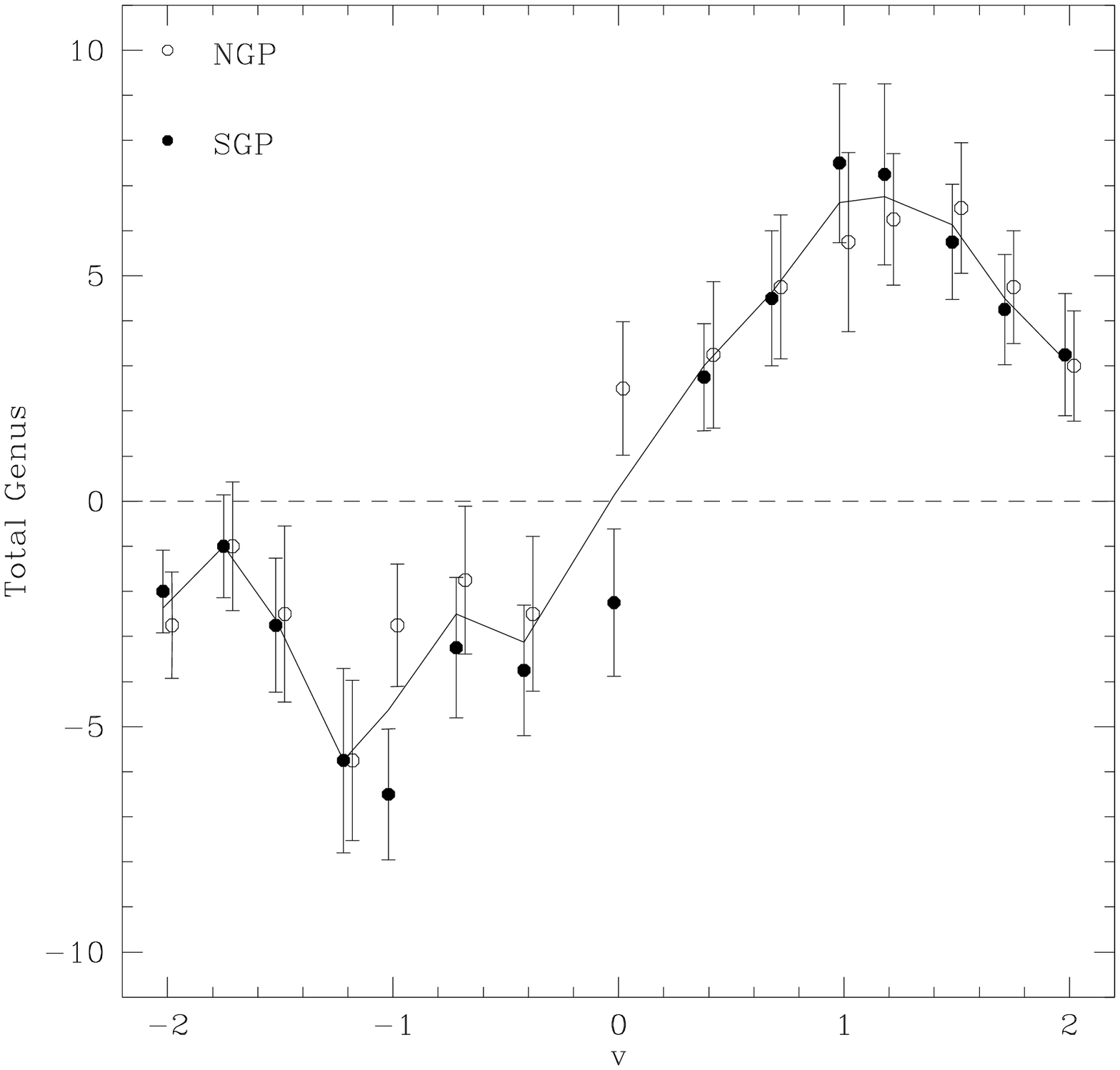}} \\
\end{tabular}
\caption{The genus of the NGP strip (open circles) and the SGP strip (filled circles) and the average genus (solid line). The points are offset slightly for clarity. A smoothing length of 10$h^{-1}$ Mpc was used in calculating the genus.}
\label{fig:2dfgenus}
\end{centering}
\end{figure}

\section{Results}
\label{sec:res}

In Figure \ref{fig:2dfgenus}, we present the genus of the 2dFGRS. We
show the genus of the NGP (open circles) and the SGP (filled circles)
as well as the average genus (solid line). The NGP and SGP genus
curves have similar shapes, although the curves cross zero at
different points. This is discussed in more detail in section
\ref{sec:meta}. The amplitude of the SGP genus curve is also slightly
higher than that of the NGP. In order to assess the significance of
these results we compare the results to simulations and use a
Mann-Whitney Rank Sum test to compare the data to the mock catalogs.

\subsection{Simulations}

To test the significance of the genus we measure from the 2dFGRS, we
compare the results to a simulation, described in section
\ref{sec:sims}. We extract 20 NGP and 20 SGP mock surveys from
the $\Lambda$CDM Hubble Volume $z=0$ simulation.

\begin{figure} 
\begin{centering}
\begin{tabular}{ccc}
{\epsfxsize=5truecm \epsfysize=5truecm \epsfbox[40 120 600 700]{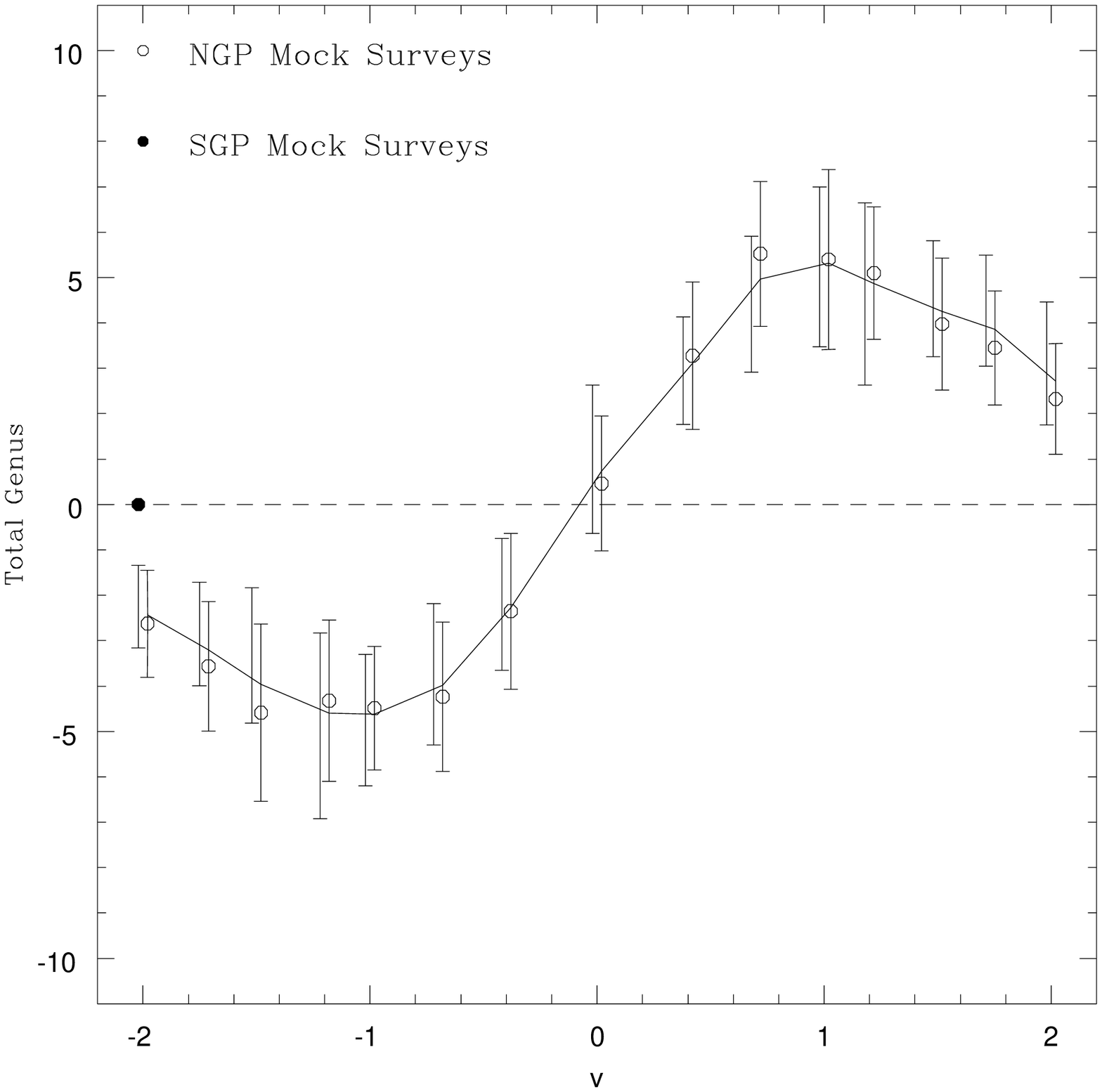}} &
{\epsfxsize=5truecm \epsfysize=5truecm \epsfbox[40 120 600 700]{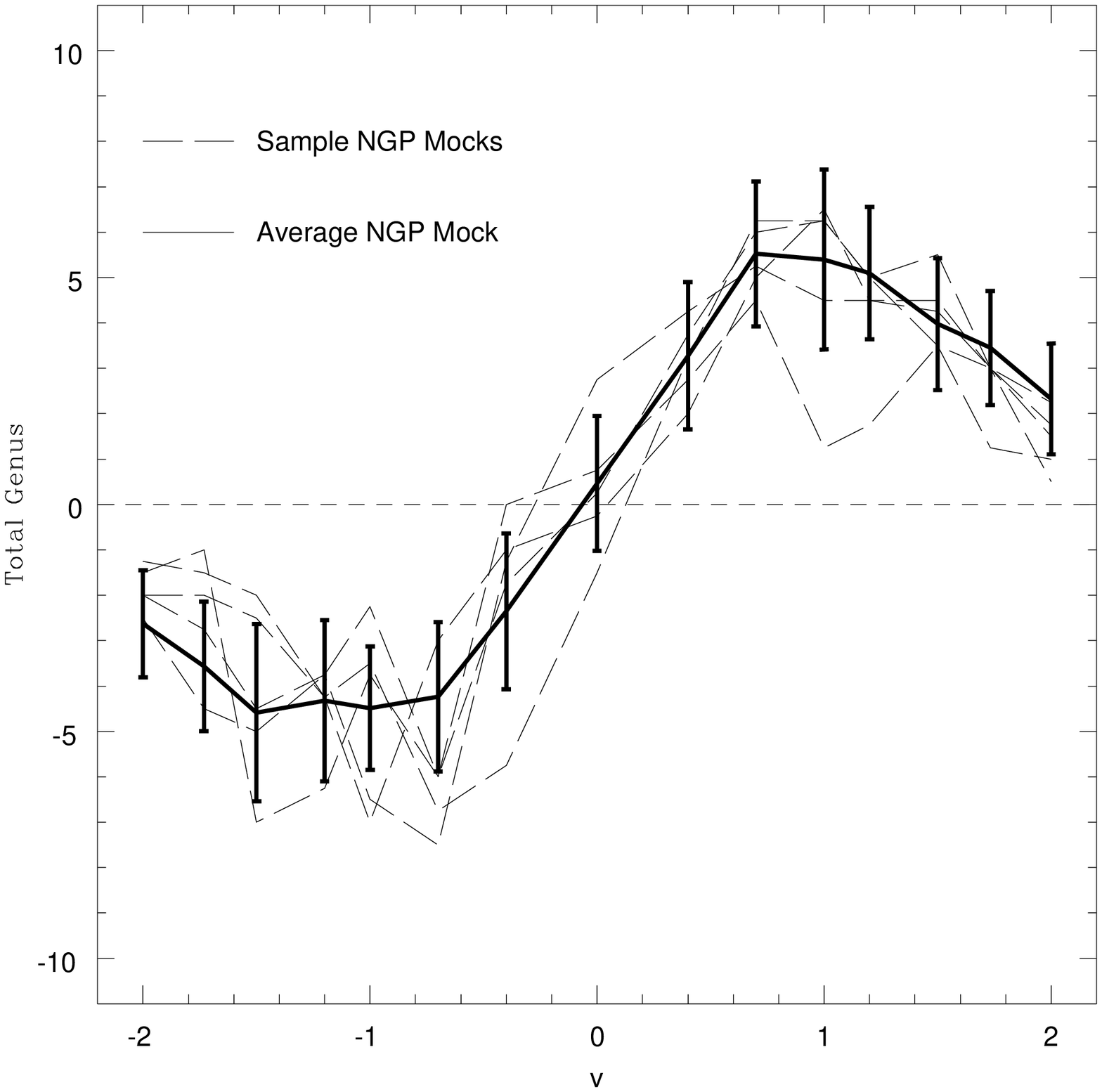}} &
{\epsfxsize=5truecm \epsfysize=5truecm \epsfbox[40 120 600 700]{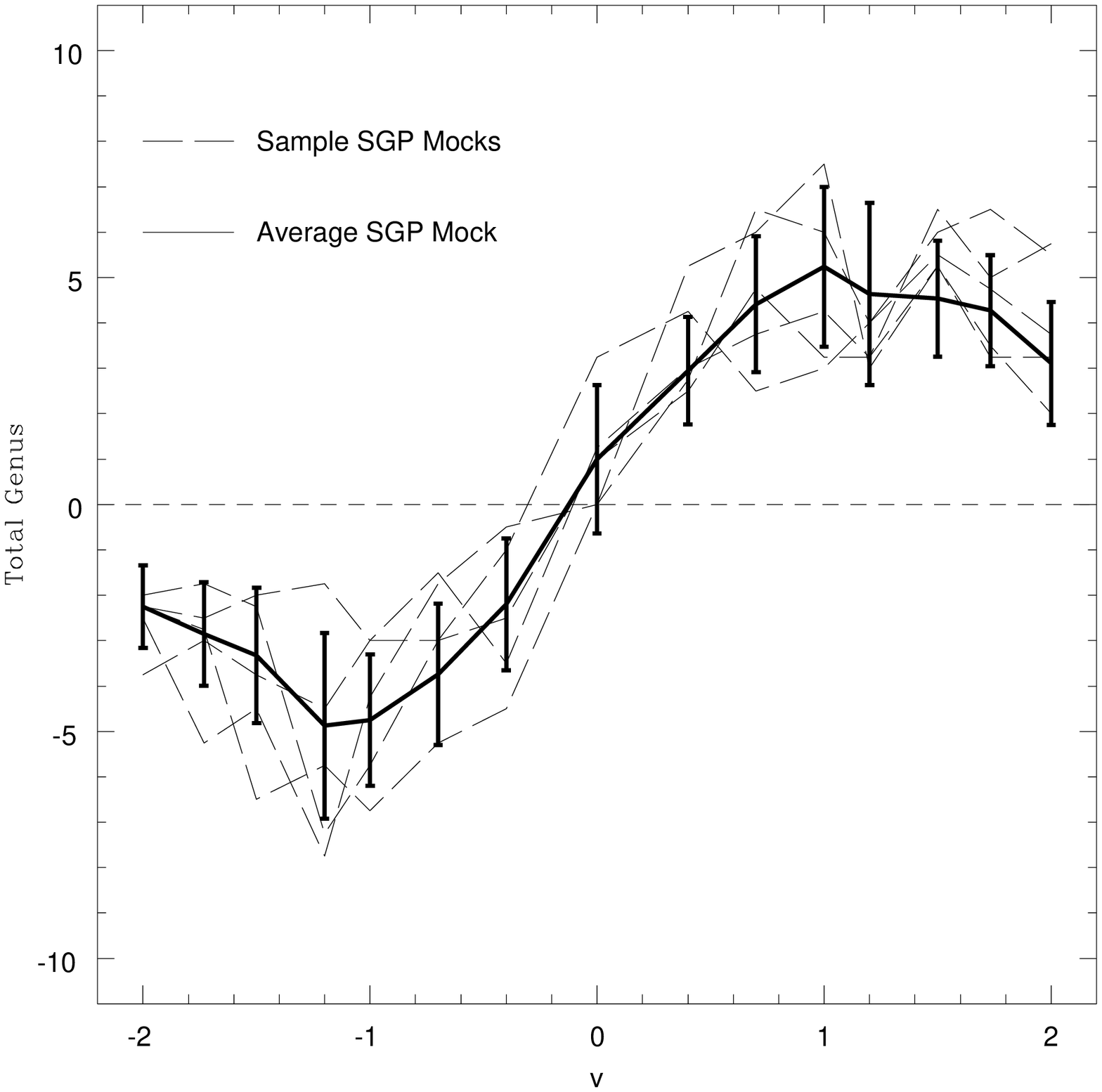}} \\
\end{tabular}
\caption{The left hand plot compares the average genus of the
$\Lambda$CDM mock NGP (open circles) and SGP (filled circles) catalogs
and the genus found by averaging the NGP and SGP together (smoothing
length 10$h^{-1}$Mpc). The points are slightly offset for clarity. The
center and right hand plots show the NGP and SGP mock catalogs with the
average genus shown by the dark line with error bars. 5 of the 20
independent realizations are shown in these figures to give an idea of
the range of genus curves that can be obtained with the current 2dFGRS
geometry. }
\label{fig:sims}
\end{centering}
\end{figure}

We measure the genus of the 20 realizations of each slice and compute
the average. We find that despite the different areas and thicknesses
of the SGP and NGP regions the genus curves are very similar, shown in
figure \ref{fig:sims}. The final SGP slice will be 5 degrees thicker
than the NGP slice but currently the data cover similar declination
ranges and thus similar genus curves are expected if the NGP and
SGP data are large enough to cover representative regions of the
universe. See Colless et al. (2001) for the masks of the public data
release.

\begin{figure} 
\begin{centering}
\begin{tabular}{c}
{\epsfxsize=6truecm \epsfysize=6truecm \epsfbox[40 120 600 700]{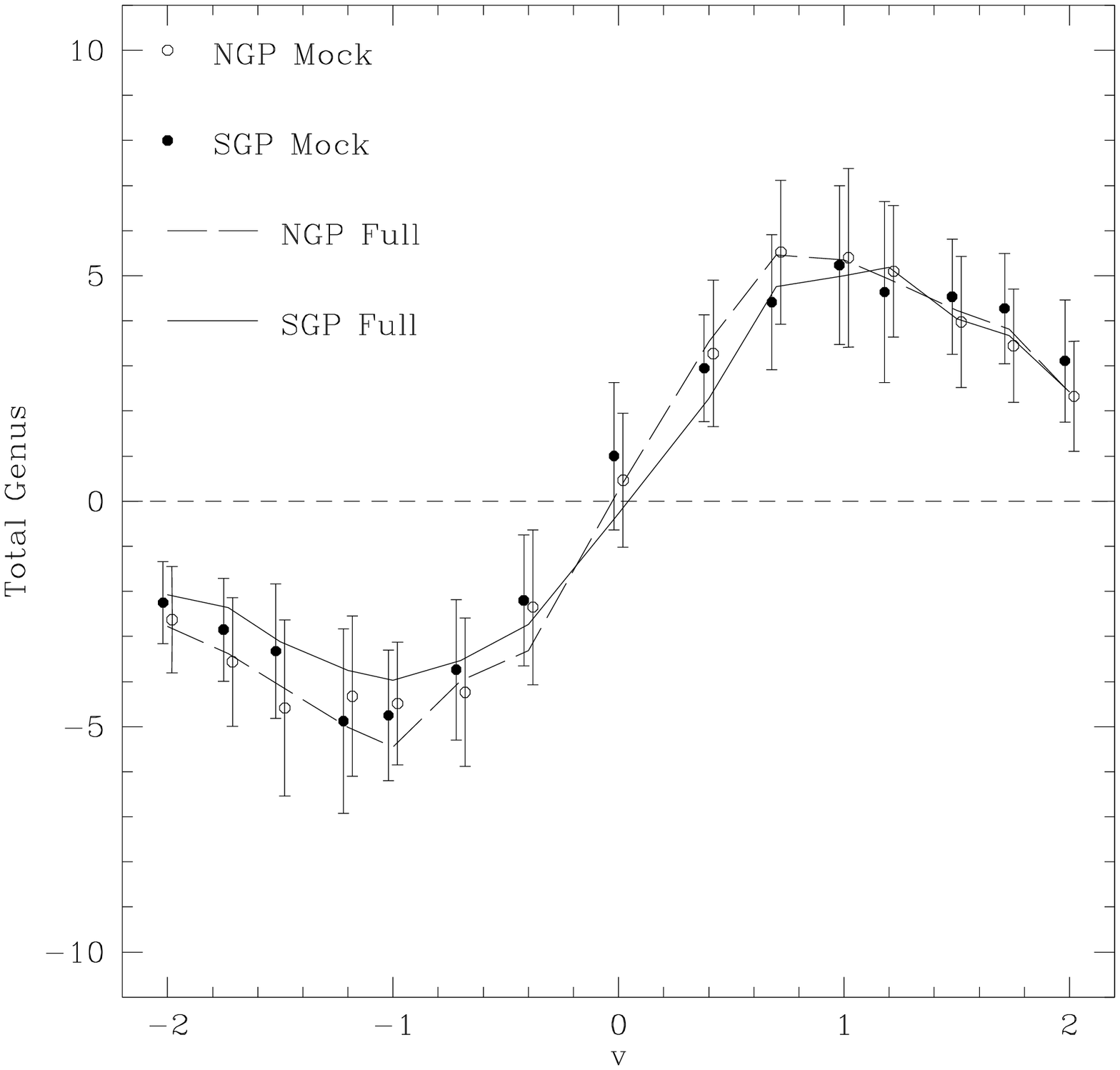}} \\
\end{tabular}
\caption{A comparison of the genus from the mock NGP and SGP
catalogs with the current incomplete geometry imprinted onto them
(open points - NGP, filled points - SGP) to the NGP and SGP genus
curves we predict for the completed regions (dashed lines - NGP, solid
lines - SGP). All curves are for Gaussian smoothing length $\lambda=10h^{-1}$Mpc.}
\label{fig:geom}
\end{centering}
\end{figure}

To test if the current geometry of the 2dFGRS significantly affects
our estimate of the genus, we compare the genus derived from the mock
catalogs to the genus found from simulation wedges that cover the
full extent of the 2dFGRS (assuming the same averaged number density
of points in the currently unobserved regions). Figure \ref{fig:geom}
shows the predicted final genus curve (solid line SGP, dashed line
NGP) compared to the mock catalog genus curve for the NGP (open
circles) and the SGP (filled circles). We see very little difference
between the two NGP curves because the current NGP data covers most of
the angle range of the expected final region. This shows that, despite
the regions in the slice where there is no data, the genus curve can
still be accurately recovered. In other words, the isolated peaks and
holes in the density distribution are adequately sampled in the
incomplete density distribution.

There is more difference between the simulated incomplete and complete
SGP curves as the final slice will cover a region that is
approximately 5 degrees thicker in declination than the current
data. The thicker slice suppresses the number of isolated high and low
density regions that can be detected because the data is effectively
smoothed in the declination direction. This gives rise to a less than
1$\sigma$ difference in the curve, suggesting the genus curve
presented here will be very similar to that from the final sample. If
the complete data covers the full 15 degree declination range in the
south, a scaling factor may be required if the genus of the NGP and
SGP are to be averaged together or a 10 degree region could be
extracted from the SGP region.

\begin{table}
\begin{centering}
\begin{tabular}{cccccccc}
Data Source & Amplitude & Rank & Significance & Shift & Rank & Significance \\ \hline
2dFGRS NGP & 8.2$\pm1.2$ & 9 & 0.42 $< 1\sigma$ & -0.20$\pm-0.09$ & 6 & 0.28 $< 1\sigma$ \\
2dFGRS SGP & 9.2$\pm1.0$ & 15 & 0.71 $< 1\sigma$ & 0.18$\pm-0.07$ & 18 & 0.86 $\sim 1\sigma$\\ \hline
2dFGRS Combined & 8.8$\pm0.85$ & 13 & 0.62 $< 1\sigma$ & -0.02$\pm-0.06$ & 12 & 0.57 $< 1\sigma$ \\ \hline
\end{tabular}
\caption{The rank of the amplitude and shift for the NGP, SGP and
average genus curves (smoothing length 10$h^{-1}$Mpc) as compared to the 20 realizations. The ranking
gives the position of the data with respect to the 20 mock catalogs,
i.e. a ranking of 9 means there are 8 mock catalogs with smaller
values and 12 with larger values of the feature we are testing.}
\label{tab:rank}
\end{centering}
\end{table}

\subsection{Error Analysis}

We use the simulations as a method for estimating the errors on the
genus curve measured from the 2dFGRS. The two-point clustering amplitude of the
$\Lambda$CDM Hubble Volume simulation is close to that of present day,
optically selected galaxies (see Hoyle et al. 1999). In figure
\ref{fig:simcomp}, 
we compare the average simulation NGP and SGP genus
curve with the curves measured from the 2dFGRS for two choices of
the smoothing length. We find that the two
curves have approximately the same shape and amplitude and therefore
it seems reasonable to use the errors from the simulation as estimates
for the errors in the data. 

The individual genus points are not independent, thus it is important
to evaluate the full covariance matrix of the measurement. We average
the mock NGP and SGP genus curves together to create mock average
genus curves. From these, we calculate the covariance matrix. This is
shown graphically in figure \ref{fig:covar}, where for clarity the
value of the covariance in each bin has been normalized according to
$C^{\prime}_{i,j} = C_{i,j}/\sqrt{(C_{i,i} C_{j,j})}$. The covariance
matrix is strongest down the diagonal elements, as expected. Around
the peaks of the genus curve, near $\nu =$ -1 and 1, there is strong
covariance between neighboring points.  This covariance makes
physical sense; the maximum amplitude of the genus curve depends on
the finite number of isolated clusters or voids that lie within the
survey region.  The peaks of the genus curve are
anti-correlated with each other. This result might be explained by the
anti-correlation between the physical locations of clusters and voids;
a region of space with a few extra clusters is less likely to have 
an excess of large voids within it.

\subsection{Amplitude and Shift of the Genus}
\label{sec:meta}

To quantify departures away from the random phase genus curve, we can
compute two statistics, termed Genus Meta-Statistics (see Vogeley et
al. 1994 for an application in three dimensions). These are the
amplitude of the genus curve and the shift of the genus curve
(i.e. where the curve has a genus value of zero [a Gaussian random
field has zero genus at the median density of $\nu = 0$]). We use these
two statistics to compare the measured genus curve with that of a
Gaussian random field and an $\Lambda$CDM N-body simulation.  The
amplitude contains information about the slope of the power spectrum;
the more power at large scales, the lower the amplitude of the genus
curve. The amplitude of the genus curve also contains information
about the non-linear clustering. A Gaussian random field that has not
been allowed to evolve under gravity will have a higher genus
amplitude than a field with the same power spectrum that has undergone
non-linear evolution. We discuss this effect in more detail in below.

To measure the amplitude, we simply find the best fit Gaussian random
phase curve (equation \ref{eq:genus}) to the data to minimize
$\chi^2$. We could use the full covariance matrix in this calculation
to prevent the result being biased by high or low points that are
highly correlated. However, an accurate estimation of the covariance
is hard to obtain as it depends on very high order moments of the
galaxy density field. Even if the true clustering of galaxies is 
well described by the
$\Lambda$CDM model, 20 mock surveys may be insufficient to accurately
estimate the covariance.

We fit the amplitude of the mock catalog genus curves to see how
strongly the off-diagonal covariance matrix points affect the fitting
of the amplitude. The amplitudes are slightly higher using only the
diagonal elements as the points around the peaks are most strongly
correlated, although we find that the amplitude of the fits to the
simulations are consistent whether we only use the diagonal elements
or the full covariance matrix. However, when we fit the 2dFGRS data,
we find very different results with and without the off diagonal
matrix elements. In the case of the NGP the best fit amplitude when we
use the full covariance matrix is 4.2$\pm1.1$ , which is clearly too
low. This value is obtained due to points with negative $\nu$ which do
not fit a Gaussian shape. When we use only the diagonal elements we
find a value of 8.2$\pm1.2$ for the NGP sample. Visually this looks
consistent with the data. The SGP does not have the same problem; we
find the result is 9.2$\pm1.0$ using the diagonal elements only and
8.8$\pm0.9$ using the full covariance matrix. We therefore just use
the diagonal elements when fitting the amplitude of the data but we
add a word of caution that these may be slightly high by $\sim 10\%$
due to the covariance between the points around the peaks of the genus
curve. These values are given in table \ref{tab:rank}.

If the measured genus crosses zero at values less than zero then the
data has a meatball topology, i.e. the isolated clusters are more
emphasized then the low density regions, if the shift is to the right
then the data has a bubble topology, i.e. there is an excess of
isolated voids compared to the high density regions. Biased CDM models
tend to display a more meatball like topology whereas hot dark matter
models tend to have a bubble topology (Melott et al. 1989). We
determine the zero point shift by simply looking at where the genus
curves have zero value. This value can be read off the genus plots
directly.

We compare the shift and amplitude values that we obtain from the 2dFGRS to
the values that we estimate from the 20 mock catalogs. We perform a
nonparametric, Mann-Whitney rank sum test to examine whether results
from the data are consistent with the simulation. The rank of the
various meta statistics as measured from the data is given in table
\ref{tab:rank}. A rank of 8, for example, means there are 7 mock
catalogs with smaller amplitude or lower zero point crossing than
the data and thus 13 mock catalogs with larger values.

The NGP strip has a slightly lower amplitude than the SGP strip. The
points near $\nu=-1$ have a lower amplitude in the NGP strip than in the
SGP strip. This is probably the main source of the difference. The
difference is around 1$\sigma$ only.

The NGP data also display a slight meatball topology whereas the SGP
displays a bubble topology. On average, the difference is very small
and consistent with the genus of a Gaussian random distribution.

\begin{figure} 
\begin{centering}
\begin{tabular}{ccc}
{\epsfxsize=5truecm \epsfysize=5truecm \epsfbox[40 120 600 700]{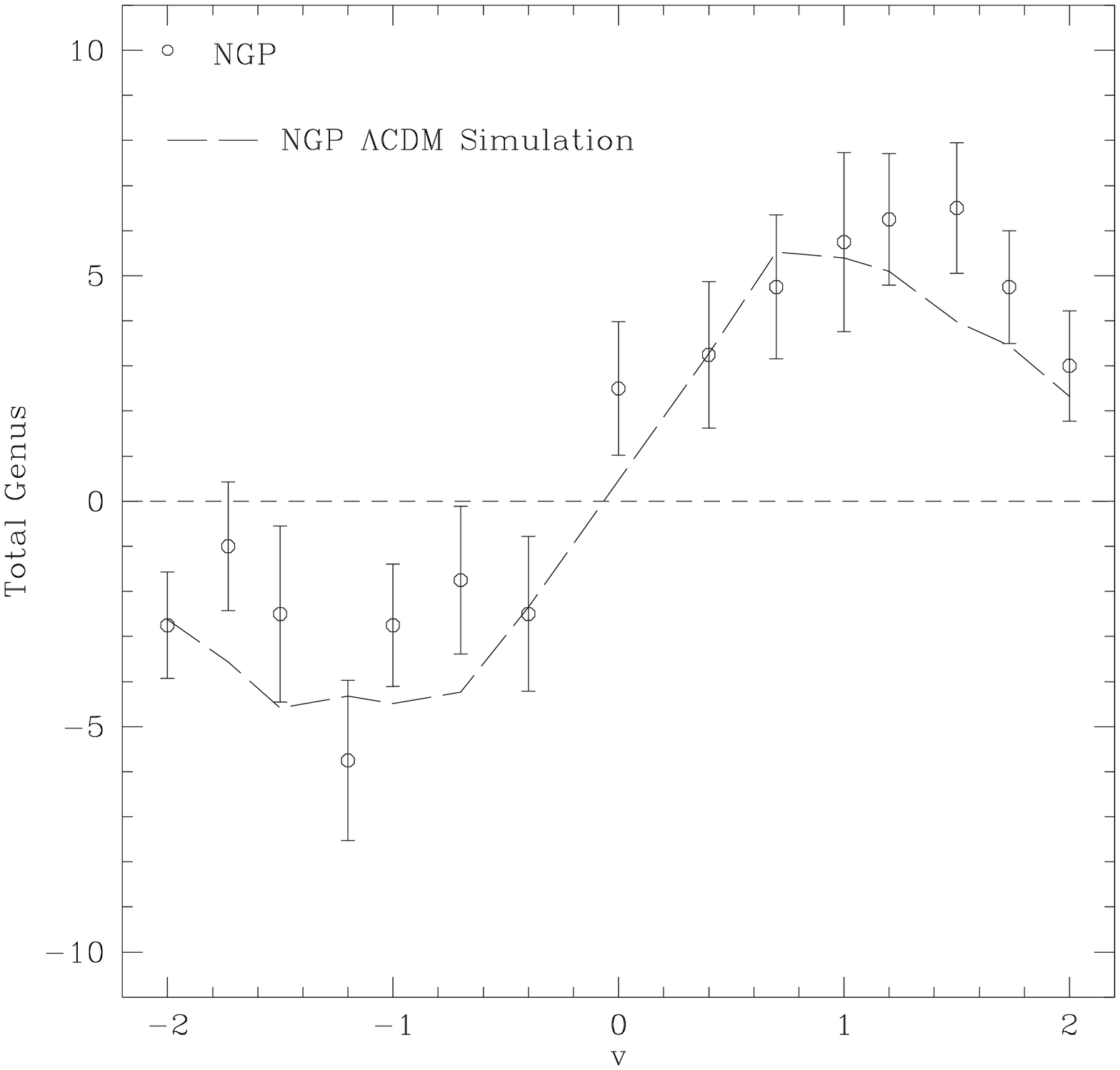}} &
{\epsfxsize=5truecm \epsfysize=5truecm \epsfbox[40 120 600 700]{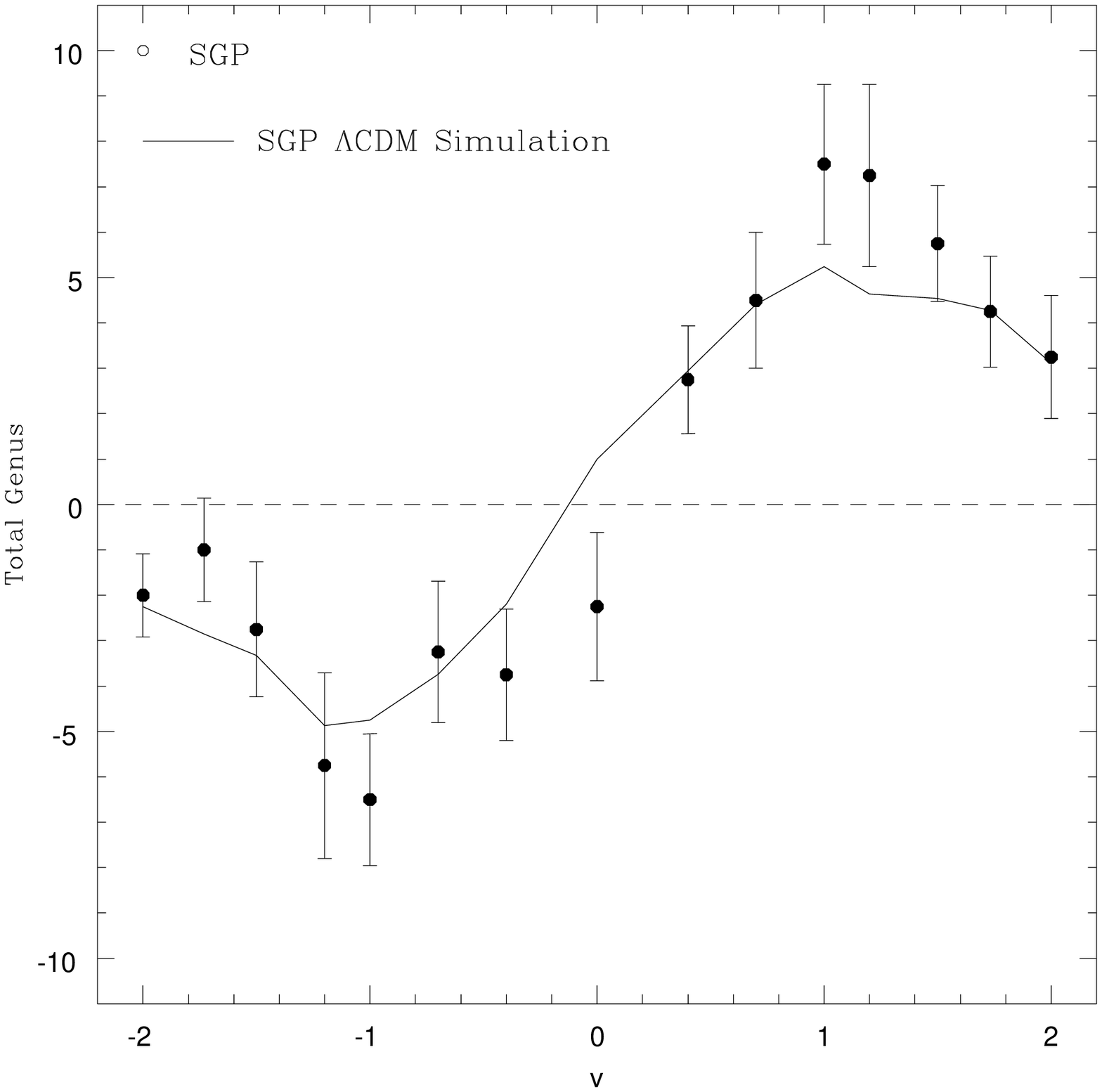}} &
{\epsfxsize=5truecm \epsfysize=5truecm \epsfbox[40 120 600 700]{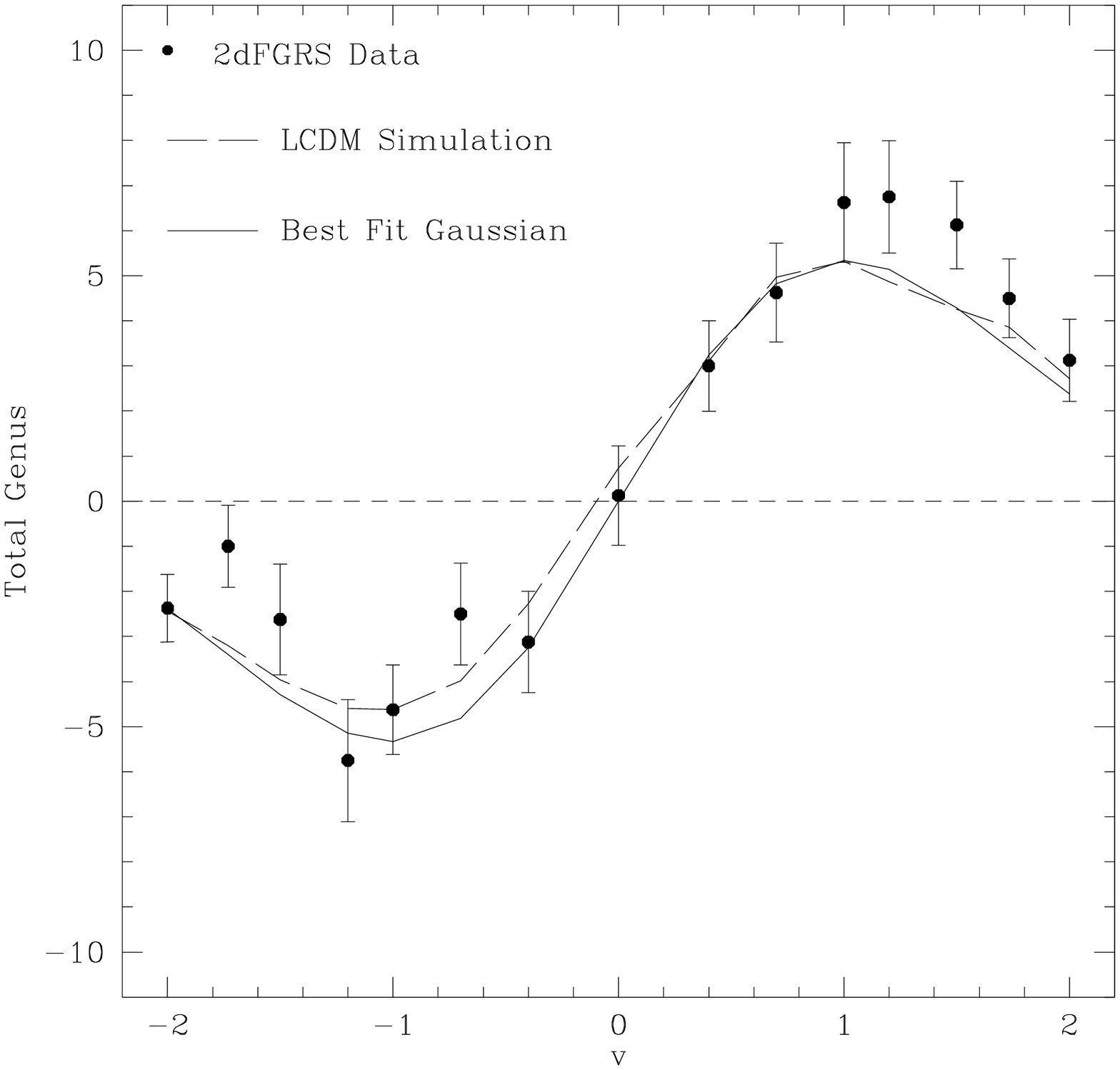}} \\
\end{tabular}
\caption{A comparison of the measured 2dF genus compared to the mock
catalog $\Lambda$CDM simulation genus curve (smoothing length 10$h^{-1}$Mpc). On the left we show the
results for the NGP. In the center we show the results for the SGP. On
the right, we compare the average genus of the 2dFGRS with the genus
of the $\Lambda$CDM simulation (dashed line) and a best fit random
phase curve (solid line) where the amplitude has been chosen to
minimize $\chi^2$.}
\label{fig:simcomp}
\end{centering}
\end{figure}

\begin{figure} 
\begin{centering}
\begin{tabular}{c}
{\epsfxsize=6truecm \epsfysize=2truecm \epsfbox[50 200 600 340]{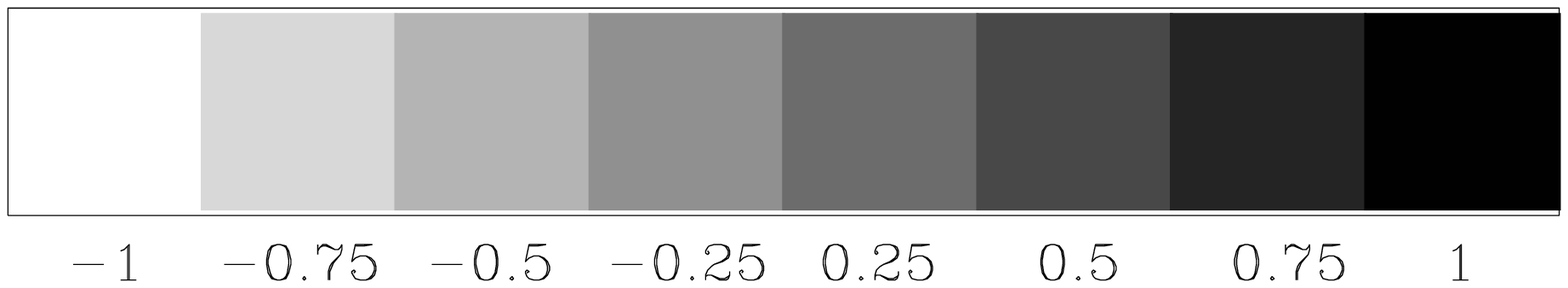}} \\ 
{\epsfxsize=6truecm \epsfysize=6truecm \epsfbox[40 120 600 700]{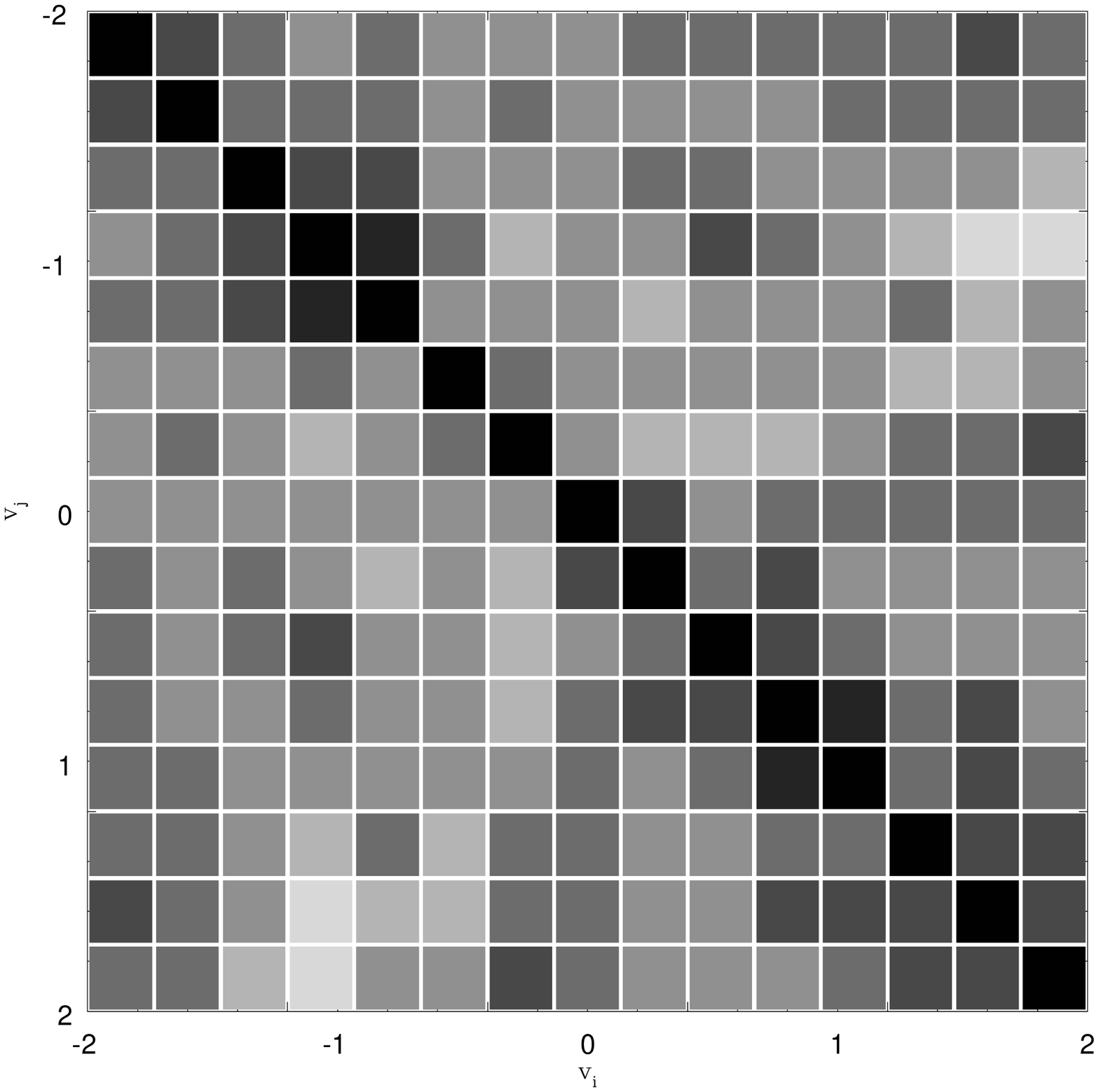}} \\
\end{tabular}
\caption{The normalized covariance matrix of genus estimates $G(\nu_i)$. 
The diagonal
elements show the variance. The off diagonal elements show the amount
of correlation between different points. We normalize the covariance
in each grid cell by $C^{\prime}_{i,j} = C_{i,j}/ (\sqrt{C_{i,i}}
\sqrt{C_{j,j}})$. The darkest shadings show where the covariance
between points is the strongest. The lightest shadings show where
points are strongly anti-correlated. Mid grey shadings show where
there is no correlation (either positive or negative) between
points. The values of $C^{\prime}_{i,j}$ shown by the different
shadings are given in the legend above the covariance matrix.}
\label{fig:covar}
\end{centering}
\end{figure}

We compare the $\chi^2$ values obtained when the data is fit by the
best fit Gaussian curve and by the $\Lambda$CDM simulation.  The
curves are plotted in the right-most panel of figure
\ref{fig:simcomp}.  We find that the minimum $\chi^2$ value from the
Gaussian curve is 25.5 where as the value of $\chi^2$ we obtain using
the simulation is 30.2. We have 15 data points which suggests a
$\chi^2$ per degree of freedom of order 2. However, the covariance
matrix tells us that adjacent points are highly correlated, so this
naive $\chi^2$ is not appropriate. Examination of the figure 5c shows
that the fits are not that bad; the significant discrepancy near
$\nu=+1$ is caused by three or four highly-correlated points that lie
above the fitted curves.  This departure suggests an excess of
clusters in this data sample relative to a Gaussian distribution or
the $\Lambda$CDM model but the excess is actually rather small when we
consider the physical significance of this peak in the genus
curve. Near $\nu=+1$ we measure the number of isolated clusters in the
sample and obtain $G(1)\sim 7$.  If the number of such structures
within a region of this size obeys Poisson statistics, then this peak
value has uncertainty of roughly $\pm \sqrt{7}$. Because peaks of a
Gaussian random field are positively correlated, the fluctuations of
the number of isolated clusters might be larger and the statistical
significance of the apparent excess of clusters in the 2dFGRS may be
quite smaller than the naive $\chi^2$ analysis suggests.

\begin{figure} 
\begin{centering}
\begin{tabular}{cc}
{\epsfxsize=5truecm \epsfysize=5truecm \epsfbox[40 120 600 700]{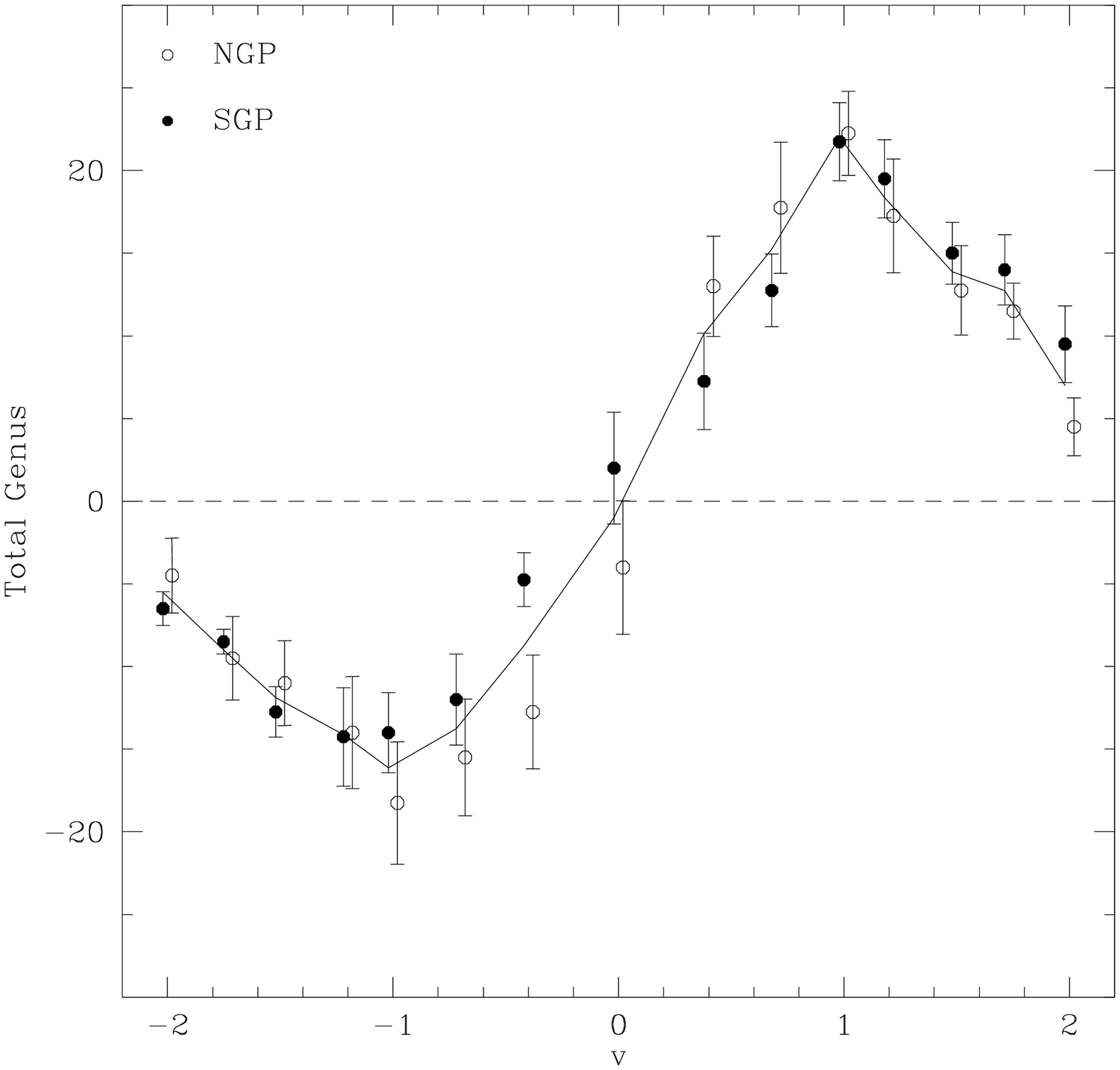}} &
{\epsfxsize=5truecm \epsfysize=5truecm \epsfbox[40 120 600 700]{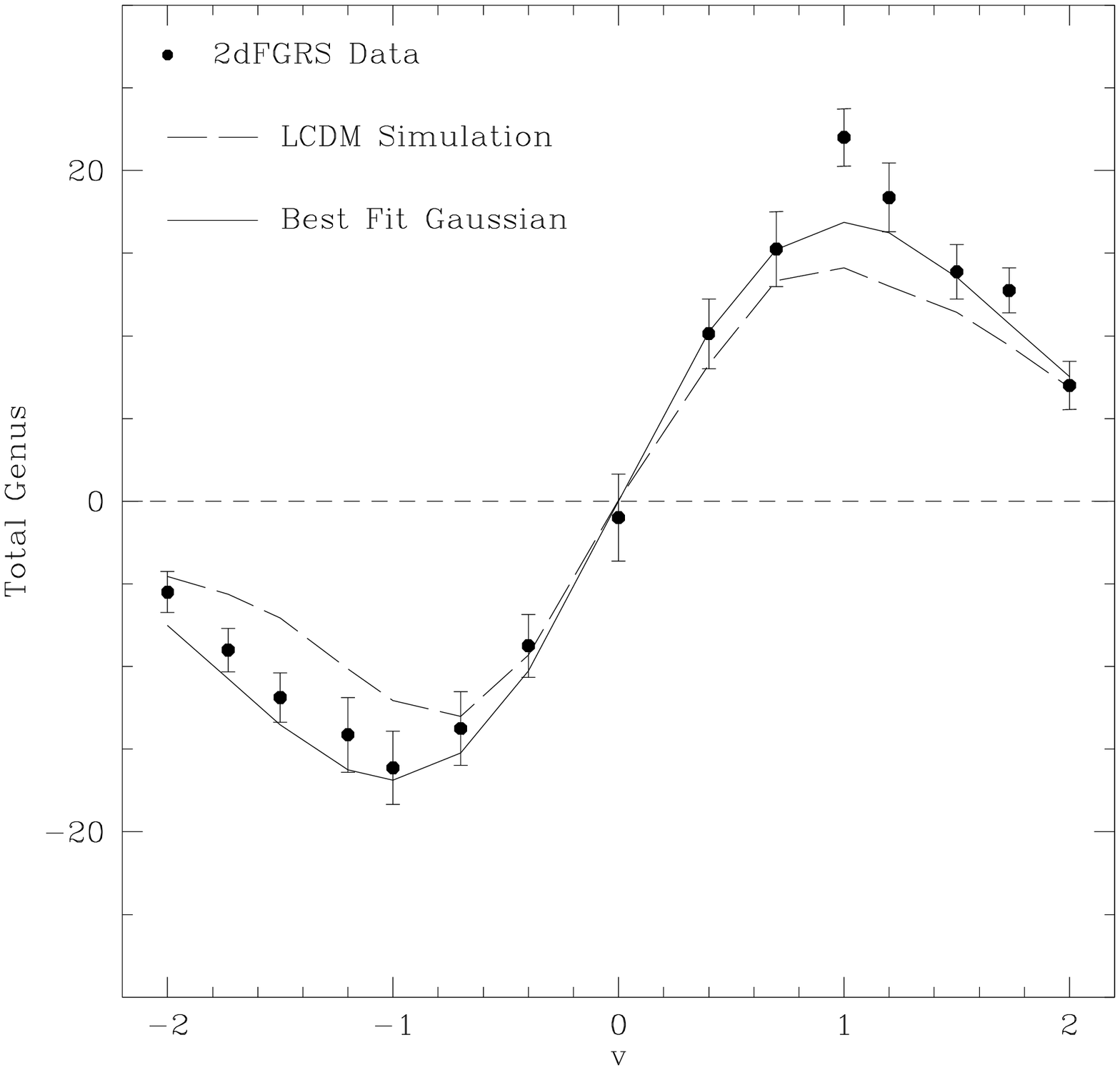}} \\
\end{tabular}
\caption{A comparison of the measured 2dFGRS genus compared to the
mock catalog $\Lambda$CDM simulation genus curve calculated using a
smoothing length of 5$h^{-1}$Mpc. On the left we show the results for
the NGP, the SGP and the average genus curve of the 2dFGRS. On the
right, we compare the average genus of the 2dFGRS calculated using a
smoothing length of 5$h^{-1}$Mpc with the genus of the $\Lambda$CDM
simulation (dashed line) and a best fit random phase curve (solid
line) where the amplitude has been chosen to minimize $\chi^2$.}
\label{fig:smooth5}
\end{centering}
\end{figure}

To examine whether the amplitude and shifts are dependent on the
choice of the smoothing length, we also compute the genus curves
for the 2dFGRS data using a smoothing of 5$h^{-1}$Mpc, using the
identical volume-limited sample of the 2dFGRS. Note that 5$h^{-1}$Mpc is
the smallest smoothing length for which both the correlation length of
galaxies and the average galaxy-galaxy separation are smaller or
comparable to the smoothing length.  The genus curves for
$\lambda=5h^{-1}$Mpc are shown in figure \ref{fig:smooth5}. We find
again that the NGP curve has a slight meatball shift, whereas
the SGP displays a slight bubble shift. There is better agreement
between the amplitudes of the two curves than for
$\lambda=10h^{-1}$Mpc, although in the 10$h^{-1}$Mpc case any
difference between the amplitudes was less than 1$\sigma$.  The
difference between the genus curve of the data and the $\Lambda$CDM
simulation using the smaller smoothing length is slightly larger. This
difference is because we are probing smaller scales at which the
clustering of the $\Lambda$CDM simulation does not match the
clustering of present day, optically selected galaxies as well (see,
for example, Hoyle et al. 1999). We compare the genus curve to a
random phase curve, where the amplitude has been selected to minimize
$\chi^2$. Again, we find that there is an excess of clusters over
voids in the 2dFGRS data. Therefore we conclude that the results we
present above are not dependent on the choice of the smoothing length.

\begin{figure} 
\begin{centering}
\begin{tabular}{c}
{\epsfxsize=6truecm \epsfysize=6truecm \epsfbox[40 120 600 700]{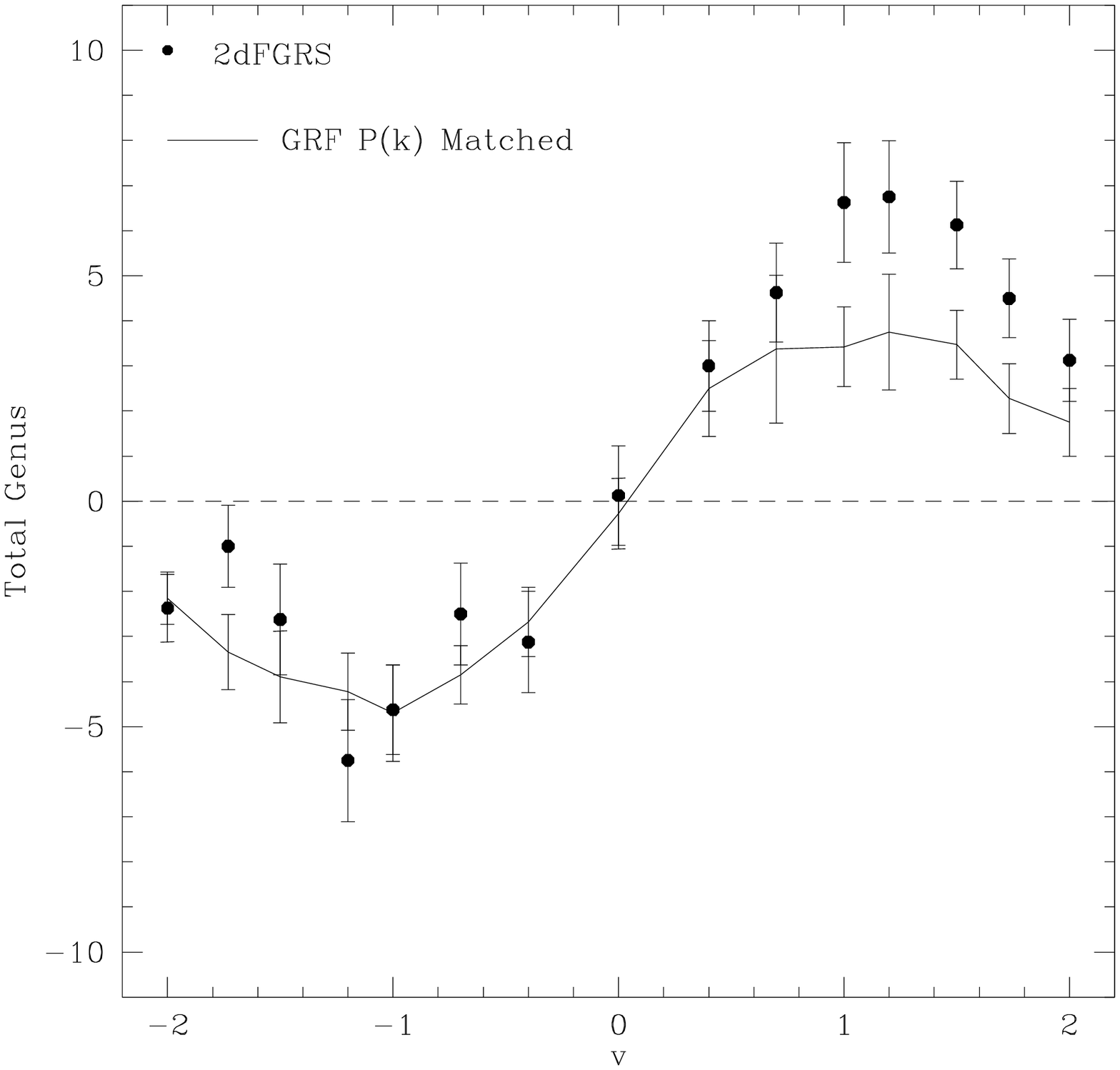}} \\
\end{tabular}
\caption{The 2dFGRS genus curve (smoothing length 10$h^{-1}$Mpc) compared to a genus curve taken from
10 realizations of a Gaussian Random Field with the power spectrum
matching that of the 2dFGRS. We detect an excess of isolated clusters
in the 2dFGRS genus curve.}
\label{fig:grfgenus}
\end{centering}
\end{figure}

\subsection{Gaussian Random Field Comparison}

As a final comparison, we compare the genus of the 2dFGRS with the
genus of a Gaussian Random Field (GRF) that has the same power
spectrum as the 2dFGRS. The power spectrum of the 2dFGRS has been
measured by Percival et al. (2001) but is only presented in a form
where it is divided by a linear theory power spectrum with values of
$\Omega_{\rm m} h = 0.2, \Omega_{\rm b}/\Omega_{\rm m} = 0$ with $n$=1
and $\sigma_8$=1. We use CMBFAST to construct such a power spectrum,
assuming that $h$=0.7 and that $\Omega = \Omega_{\rm m} +
\Omega_{\Lambda}$=1. We multiply the points presented in Percival et
al. to obtain an estimate of the 2dFGRS power spectrum. We also
measure the power spectrum from the two NGP and SGP volume-limited
samples we have constructed here to check consistency and find that
the three power spectra have approximately the same shape. On scales
larger than 10$h^{-1}$Mpc and up to 400$h^{-1}$Mpc, the 2dFGRS P(k) is
fairly similar in shape to the CMBFAST power spectrum. To test how
sensitive the genus curve is to the exact features in the power
spectrum curve, we compare the genus curves we obtain using a power
law with P$\propto k^{-2}$, the CMBFAST power spectrum and the
estimate of the 2dFGRS power spectrum we obtain here and find little
difference between the CMBFAST and 2dFGRS genus curves and only small
differences between these genus estimates and ones created with the
power law input. We construct the GRF on a 256$^3$ array by assigning
values to the real and imaginary parts of each Fourier mode which are
Gaussian deviates. We then imprint the shape of the NGP and SGP
regions onto the grid before we smooth the density field and calculate
the genus, exactly as described above.

We compare the genus of the 2dFGRS with that of a GRF
with the same power spectrum to see if the non-linear clustering
that appears in the data and in the simulations affects the amplitude
of the 2D genus. In previous papers that examine the genus in 3D,
significant damping of the amplitude of the data genus curve has been
found as compared to the genus curve of a GRF. Vogeley et al. (1994)
found that for smoothing scales of 6$h^{-1}$Mpc that a GRF had a
2$\sigma$ higher amplitude than the genus curve of the CfA survey. On
larger smoothing scales this difference was reduced and at
16$h^{-1}$Mpc, the two curves had consistent amplitude.

We find in 2D that we do not see the same damping that has previously been
seen in the 3D genus. For values of $\nu < 0$, the amplitude of the GRF genus
curve is fairly similar to that of the 2dFGRS but on positive $\nu$
scales, the amplitude is lower than the 2dFGRS. On these large scales,
the data is definitely showing an excess of overdensities, as shown earlier in
figure \ref{fig:simcomp}. The best fitting Gaussian curve has a
comparable amplitude to the underdense part of the genus curve but a
lower amplitude than the overdense part of the genus curve.

Several factors could account for the apparent discrepancy between the
2 and 3D genus amplitude ratios. It may be that our smoothing length
of 10$h^{-1}$ is close enough to the linear regime that no amplitude
difference due to non-linear clustering may be seen.  Note that our
10$h^{-1}$Mpc smoothing length would be $\lambda =14.14h^{-1}$Mpc if
we used the smoothing length definition adopted in in older genus
papers. Vogeley et al. (1994) found no effect in 3D for
$\lambda=16h^{-1}$Mpc.  It may also be that the non-linear clustering
has a smaller effect on the 2D genus, where data are projected and
thereby smoothed over the width of the survey, thus making the
effective smoothing length larger.  However, we repeated this analysis
using a 2D smoothing length of only 5$h^{-1}$Mpc and found similar
results.  Because an excess of clusters may reflect biasing of the
galaxies with respect to the mass distribution, it will be interesting
to repeat this analysis with the final data set to examine whether the
completed sample exhibits this amplitude excess in the genus curve at
high density thresholds.  If these partial data in the 2dFGRS strips
were slightly biased toward high density regions, perhaps because
denser fields were given targeting priority over lower density
regions during early observations, then the complete sample might 
lack such a feature.

\section{Conclusions}
\label{sec:conc}

We measure the genus from a 100,000 galaxy redshift sample of the
2dFGRS, the largest redshift sample publicly available to date.  We
find that the genus curves of the NGP and SGP are slightly different,
with the NGP region displaying an excess of isolated clusters and the
SGP region displays a slight excess of isolated voids. These shifts
are only significant at the 1$\sigma$ level when we compare the genus
curves to mock catalogs drawn from a $\Lambda$CDM simulation. The results 
are qualitatively the same for two choices of the smoothing length. We
average the NGP and SGP data and find that the average genus curve is
consistent with the $\Lambda$CDM simulation. We find that the average
2dFGRS genus curve reveals a smaller number of isolated voids than
clusters in the Survey, although the points in the genus curve that
show this slight excess of clusters are highly correlated.

We demonstrate that in spite of the current incompleteness in
the geometry, the genus can be accurately recovered. No bias is
introduced into the shape of the genus curve by the current angular
coverage.

We compare the 2dFGRS genus curve with that for a GRF with the same
power spectrum. We do not find any significant amplitude depression in
the amplitude of the 2dFGRS genus curve due to non-linear
clustering. This may be due to the large smoothing length, may be a
feature of the 2D genus, or may indicate a real excess in the number
of isolated clusters in the the 2dFGRS
that is not predicted directly from the power
spectrum alone. Colley (1997) found that the 2D genus of the Las
Campanas Redshift Survey agrees well with a Gaussian random field. A
larger sample of galaxies will reveal whether there really is an
excess of over dense regions, or whether this is a feature of the
observing strategy applied so far.

\section*{Acknowledgments} 
MSV acknowledges support from NSF grant AST-0071201 and the John
Templeton Foundation. JRG acknowledges the support from NSF grant AST-9900772. 
We thank David Weinberg for providing the CONTOUR2D code. 
We also thank the 2dFGRS team for their tremendous efforts in obtaining
and releasing this data set and for providing the related analysis
tools.

\end{document}